\documentclass{ieeeaccess}
\usepackage{cite}
\usepackage{amsmath,amssymb,amsfonts}
\usepackage{algorithmic}
\usepackage{graphicx}
\usepackage{textcomp}
\usepackage{booktabs}

\newcommand{\antcom}[1]{\textcolor{black}{#1}}

\newcommand{\pp}[1]{\textcolor{black}{#1}}
\def\BibTeX{{\rm B\kern-.05em{\sc i\kern-.025em b}\kern-.08em
    T\kern-.1667em\lower.7ex\hbox{E}\kern-.125emX}}
\begin{document}
\history{Date of publication xxxx 00, 0000, date of current version xxxx 00, 0000.}
\doi{xxxxxx}

\title{LEO Small-Satellite Constellations for 5G and Beyond-5G Communications}
\author{\uppercase{Israel Leyva-Mayorga}\authorrefmark{1}, \IEEEmembership{Member, IEEE}, \uppercase{Beatriz~Soret}\authorrefmark{1}, \IEEEmembership{Member, IEEE},
 \uppercase{Maik R{\"o}per}\authorrefmark{2}, \IEEEmembership{Student Member, IEEE},
\uppercase{Dirk W{\"u}bben}\authorrefmark{2}, \IEEEmembership{Senior Member, IEEE}, \uppercase{Bho Matthiesen}\authorrefmark{2},
\IEEEmembership{Member, IEEE}, \uppercase{Armin Dekorsy}\authorrefmark{2}, \IEEEmembership{Senior Member, IEEE}, \uppercase{and Petar Popovski}\authorrefmark{1,2}, \IEEEmembership{Fellow, IEEE}}
\address[1]{Department of Electronic Systems, Aalborg University, 9220, Aalborg, Denmark.}
\address[2]{Department of Communications Engineering, University of Bremen, Germany.}
\tfootnote{This work has been supported in part by the European Research Council
(Horizon 2020 ERC Consolidator Grant Nr. 648382 WILLOW), by the European Regional Development Fund (ERDF) under grant LURAFO2012A, and by the German Research Foundation (DFG) under Germany's Excellence Strategy (EXC 2077 at University of Bremen, University Allowance)}

\markboth
{Leyva-Mayorga \headeretal: LEO Small-Satellite Constellations for 5G and B5G}
{Leyva-Mayorga \headeretal: LEO Small-Satellite Constellations for 5G and B5G}

\corresp{Corresponding author: Israel Leyva-Mayorga (e-mail: ilm@es.aau.dk).}

\begin{abstract}
The next frontier towards truly ubiquitous connectivity is the use of Low Earth Orbit (LEO) small-satellite constellations to support 5G and 
Beyond-5G (B5G) networks. Besides enhanced mobile broadband (eMBB) and massive machine-type communications (mMTC), LEO constellations can support ultra-reliable communications (URC) with relaxed latency requirements of a few tens of milliseconds. Small-satellite impairments and the use of low orbits pose major challenges to the design and performance of these networks, but also open new innovation opportunities. This paper provides a comprehensive overview of the physical and logical links, along with the essential architectural and technological components that enable the full integration of LEO constellations into 5G and B5G systems. Furthermore, we characterize and compare each physical link category and explore novel techniques to maximize the achievable data rates.
\end{abstract}


\titlepgskip=-15pt

\maketitle

\section{Introduction}
\label{sec:introduction}
\PARstart{C}{onstellations} of small satellites flying in Low Earth Orbits (LEO) and working all together as a communication network 
present an attractive solution to support and complement 5G New Radio (NR) and Beyond-5G (B5G) communications~\cite{3GPPTR38.913, 3GPPTR22.822, 3GPPTR38.821, Di2019}.

These constellations are deployed at altitudes between $500$ and $2000$~km and their integration with 5G NR will provide nearly-global coverage and support for: 1) enhanced mobile broadband (eMBB), to offer increased user data rates; 2) massive Machine-Type Communications (mMTC), to enable a wide range of Internet of Things (IoT) applications operating over vast geographical areas; and 3) Ultra-Reliable Communications (URC), to provide one-way latency guarantees in the order of $30$\,ms~\cite{3GPPTR22.822}, with typical $2$\,ms propagation delays between ground and LEO.

\begin{figure}[t]
    \centering
    \includegraphics{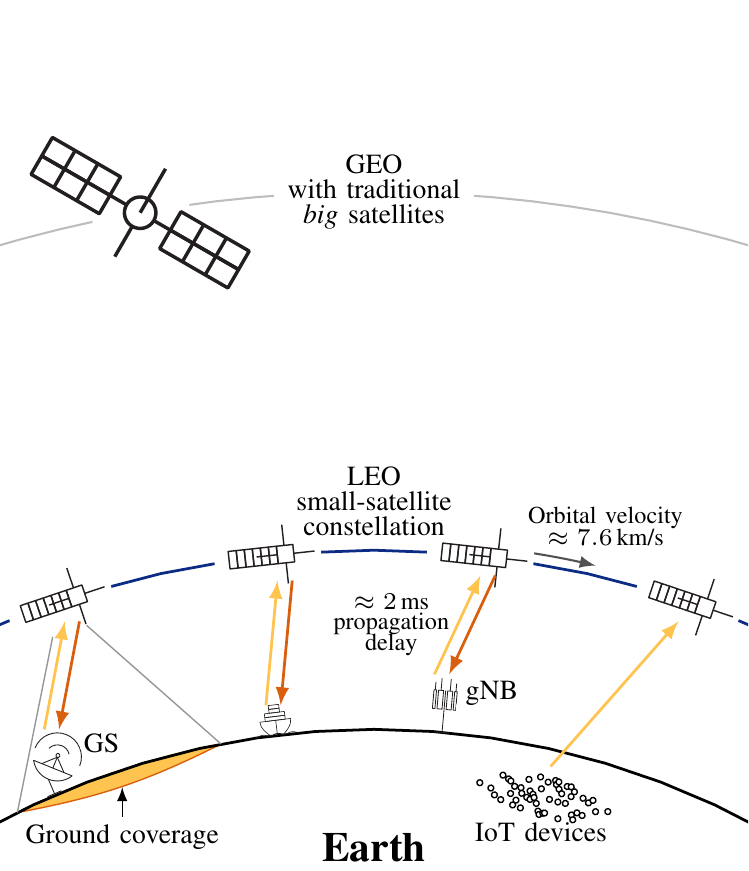}
    \caption{Overview of the unique characteristics of LEO small-satellite constellations with respect to traditional GEO satellites.}
    \label{fig:overview}
\end{figure}

Unlike Geostationary Orbits (GEO), LEO satellites  move rapidly with respect to the Earth's surface and have a small ground coverage. In particular, the ground coverage of a LEO satellite deployed at $600$~km above the Earth's surface and with a typical elevation angle of $30^\circ$ is around $0.45$\% of the Earth's surface. Moreover, due to the low altitude of deployment, LEO satellites can communicate with diverse types of ground terminals, such as dedicated ground stations (GSs), 5G gNBs, ships and other vehicles, or Internet of Things (IoT) devices. These features and elements, illustrated in Fig.~\ref{fig:overview} create the need for a relatively dense constellation 
to ensure that any ground terminal is always covered by, at least, one satellite. Therefore, global commercial deployments usually consist of more than a hundred satellites. For example, Kepler, Telesat, and Starlink constellations will consist of $140$, around $300$, and between $12000$ and $42000$ satellites, respectively~\cite{Mitry2020,Portillo2019}.

Naturally, the need for a high number of satellites introduces a constraint on their individual manufacturing and deployment cost~\cite{Kak2019}. Therefore, LEO deployments typically incorporate small or even nano satellites, with low manufacturing cost, size, and weight (i.e., under $500$~kg) when compared to the traditional \emph{big} satellites at Medium Earth Orbit (MEO) and GEO. These characteristics in turn reduce the launching costs. 
Small satellites have, however, stringent connectivity, processing, and energy constraints~\cite{Budianu2013}, exacerbated by long transmission distances and potentially large Doppler shift due to the rapid movement of the LEO space segment. 

This paper provides an overview of the unique characteristics and the challenges for LEO small-satellite constellations in the context of 5G and B5G communications. Furthermore, we analyze physical layer and radio access techniques that are not relevant for traditional GEO satellite systems but, on the other hand, are essential to unleash the full potential of dense LEO constellations.
Specifically, the 
contributions of the paper include:
\begin{itemize}
\item The description of the characteristics and communication challenges of LEO constellations, and their role in 5G and B5G.
    \item The analysis and comparison of the achievable data rates, propagation delays and Doppler shift in every physical link in the satellite constellation with parameters taken from the 3GPP technical reports~\cite{3GPPTR38.821}. 
    \item A taxonomy for the logical link types in satellite constellations. This includes the different types of data for which each of the link types can be used.
    \item The identification of the most relevant enabling technologies at the physical layer, the radio access and the radio slicing. 
    \item In the physical layer, the evaluation of the performance gains of adaptive coding and modulation and the use of multiple-input multiple-output (MIMO). In particular, we explore the benefits of distributed MIMO with a set of transmitting small-satellites flying in close formation.  
    \item In the radio access, the analysis of the benefits of resource allocation to mitigate interference with different resource types, namely orthogonal frequencies and codes. This is illustrated in terms of the effective data rates for communication between satellites. 
\end{itemize}

The rest of the paper is organized as follows. Section~\ref{sec:leoconstellations} presents an overview of LEO small-satellite constellations. Section~\ref{sec:connectivity} details and characterizes the physical and logical links. The integration of constellations in 5G and B5G is discussed in Section~\ref{sec:5Gintegration}. Furthermore, enabling technologies in the physical layer, the radio access, and the radio slicing are discussed; the benefits of a subset of technologies are also evaluated. The paper is concluded in Section~\ref{sec:conclusions}.  

\section{LEO small-satellite constellations: characteristics and challenges} \label{sec:leoconstellations}
As a starting point, we define some key terms and concepts used throughout the rest of the paper. 

A \emph{satellite constellation} is typically organized in several \emph{orbital planes}, which are groups of satellites deployed at the same altitude and inclination. A \emph{pass} is the period in which a satellite is available for communication with a particular ground position, with a typical duration of a few minutes for a LEO satellite, depending on the elevation angle and the relative position between terminals. The minimum elevation angle is typically between $10^\circ$ and $45^\circ$~\cite{3GPPTR38.821, Portillo2019}.

There are three elements present in every space mission: 1) the \emph{space segment}, in our case the satellite constellation; 2) the \emph{ground segment}, with the set of GSs, which are responsible of major control and management tasks of the space segment, plus the ground networks and other mission control centers; and 3) the \emph{user segment}, which refers to the rest of communication devices at ground level, including IoT devices, smartphones, gateways or cellular base stations.
We use the term \emph{ground terminal} to denote any communication device deployed at the ground level, encompassing the ground and user segments. 

There is a close relation between the mass, energy source, and processing and communication capabilities of a satellite. High-throughput eMBB or general-purpose space missions typically involve relatively heavy satellites (above $100$~kg). This is the case of well-known commercial missions such as Starlink~\cite{Portillo2019}. Recently, smaller nano- and pico- satellites have increased in popularity due to their low manufacturing and deployment costs, as well as the ability to support a wide range of IoT and broadband applications~\cite{Rahakrishnan2016}. Small satellites must have active lifetimes of up to five years to prevent frequent redeployment. Therefore, they usually incorporate photovoltaic panels for energy supply and to charge their batteries, which are used when sunlight is absent. This calls for an adequate balance between energy consumption and performance.

One of the main concerns in satellite communications due to the long transmission distances is latency. The total latency is a combination of processing delay, queueing delay, transmission time, and propagation delay, being the latter determined by the physical distance between source and destination. 

Interestingly, wireless communications through LEO satellites over long distances present an advantage in propagation delay with respect to terrestrial communications~\cite{Handley2018}. This is because electromagnetic waves propagate in space at the speed of light, whereas the propagation speed in optic fiber is around $1.47$~times slower. Depending on the packet lengths, data rates, and queueing delays, this advantage may lead to a lower total latency with a LEO satellite constellation than with terrestrial networks over long distances.

In terms of communication technology, both free-space optical (FSO) and traditional radio frequency (RF) are considered for communication between satellites, through \emph{inter-satellite links (ISLs)}, and with the ground terminals, through \emph{ground-to-satellite links (GSLs)}~\cite{3GPPTR38.821}. FSO links employ ultra-narrow beams to combat the increased attenuation of high carrier frequencies with distance, offering increased transmission ranges, higher data rates, and lower interference levels when compared to RF links~\cite{Kaushal2017}. FSO has been demonstrated in ground-to-satellite communication in numerous scientific missions~\cite{Kaushal2017} and several planned commercial LEO constellations, such as SpaceX, Telesat, and LeoSat, will deploy laser communication equipment for high-throughput FSO ISLs~\cite{Portillo2019}. On the downside, FSO is highly susceptible to atmospheric effects and pointing errors. Hence, communication through FSO links requires a combination of precise pointing capabilities with pre-arranged pairing, so that the antennas of the intended receiver and transmitter steer the beams accordingly. 

In contrast, RF links present wider beams that enable neighbor discovery procedures, along with the integration into terrestrial RF-based systems. For example, the communication between a ground terminal and a satellite in 5G NR is envisioned to take place either in the \mbox{S-band} around $2$\,GHz or in the Ka-band, where the downlink operates at $20$\,GHz and the uplink at $30$\,GHz~\cite{3GPPTR38.821}. Also, RF links are crucial as fallback solution if FSO communication is infeasible, for example, due to positioning and pointing errors in the space segment, traffic overload, or bad weather conditions. Hence, a hybrid RF-FSO system has great potential to enhance network flexibility and reliability. 

Finally, LEO satellites move rapidly with respect to the ground -- at up to $7.6$~km/s for an altitude of $500$~km -- but also with respect to each other in different orbital planes. This leads to two main challenges. \emph{First}, the constellations are dynamic, usually entailing slight asymmetries that are aimed at minimizing the use of propellant when avoiding physical collisions between satellites at crossing points. Therefore, dynamic rather than fixed mechanisms must be put in place to create and maintain the links. \emph{Second}, these links experience a much larger Doppler shift than those found in terrestrial systems. 

\section{Connectivity} \label{sec:connectivity}
There are three types of data traffic in a LEO constellation: 1) user data, 2) control data, and 3) telemetry and telecommand (TMTC) data. The latter are inherently different from network control data and are exchanged between the GSs and the satellites. In the downlink, telemetry parameters describing the status, configuration, and health of the payload and subsystems are transmitted. In the uplink, commands are received on board of the satellite to control mission operations and manage expendable resources, for example, propellant. Oftentimes, TMTC uses separate antennas and frequency bands.

\subsection{Physical links and performance characterization} \label{sec:phylinks}
The \emph{physical links} are the broadly classified in GSLs and ISLs. A GSL between a dedicated GS and a satellite, illustrated on the left side of Fig.~\ref{fig:overview}, is also called the feeder link. The availability of a GSL, the satellite pass, is determined by the ground coverage and the orbital velocity (see Fig.~\ref{fig:overview}). Due to the high orbital velocities of LEO satellites, these passes are short and frequent handovers between satellites are necessary to maintain connections with the ground terminals. The optimal pass corresponds to that in which, at some point in time, the satellite crosses over the observer's zenith point. 

\begin{figure}
    \centering
    \includegraphics{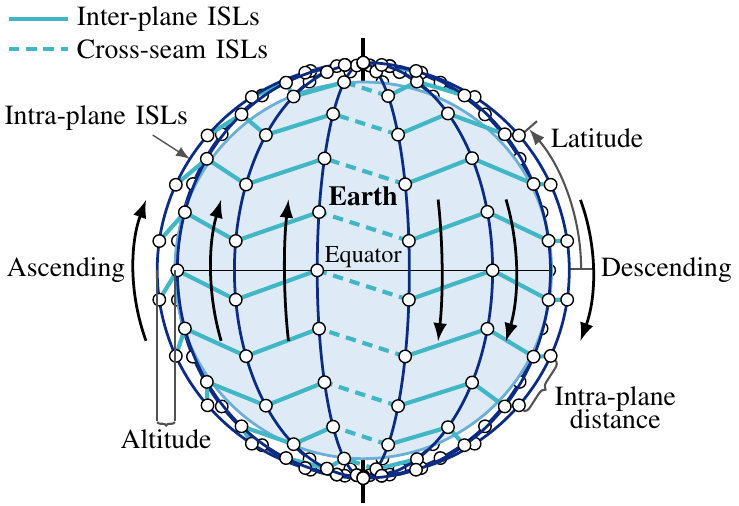}
    \caption{Diagram of a Walker star LEO constellation with the established intra- and inter-plane (including cross-seam) ISLs.}
    \label{fig:constellation}
\end{figure}

On the other hand, ISLs can be further divided into intra- (for satellites in the same orbital plane) and inter-plane ISLs (for satellites in different orbital planes). Furthermore, the ISLs between satellites in orbital planes moving in nearly-opposite directions (one ascending and on descending) are known as cross-seam ISLs. Fig.~\ref{fig:constellation} illustrates the ISLs in a typical Walker star constellation with seven orbital planes.

In this section, we characterize the GSLs and ISLs in terms of propagation delay, Doppler shift, and achievable data rates. Unless otherwise stated, the results presented were obtained by simulation with parameters taken from~\cite[Section 6]{3GPPTR38.821}. These are listed in Table~\ref{tab:param}. Note that the effective isotropic radiated power (EIRP) density is the design parameter for the satellite transmitters, whereas a fixed transmission power is defined for the ground terminals.


\begin{table}[t]
    \centering
        \caption{Parameter settings}
    \begin{tabular}{@{}lll@{}}
    \toprule
        \multicolumn{2}{@{}l}{Parameter} & Setting \\   \midrule
        \multicolumn{2}{@{}l}{\textbf{Communication}}\\
         & Carrier frequency for GSL downlink& $f_c=20$~GHz\\
         & Carrier frequency for GSL uplink and ISL& $f_c=30$~GHz\\
        & Channel bandwidth & $B=400$~MHz\\
        & EIRP density for satellites & $4$\,dBW/MHz\\
        & Antenna gains for satellites & $38.5$~dBi\\
        & Transmission power for ground terminals& $33$~dBm\\
        &Transmitter antenna gain for ground terminals & $43.2$~dBi\\
        &Receiver antenna gain for ground terminals & $39.7$~dBi\\
        &Minimum elevation angle & $30^\circ$\\
        & Atmospheric loss & $0.5$~dB\\
        & Scintillation loss & $0.3$~dB\\
        & Noise temperature & $354.81$~K\\
         \multicolumn{2}{@{}l}{\textbf{Satellite constellation}}\\
         & Number of orbital planes & $P\in\{7,12\}$\\
         & Number of satellites per orbital plane & $N\in\{20,40\}$\\
         & Altitude of orbital plane $p\in\{1,2,\dotsc,P\}$ & $600 + 10(p-1)$~km\\
         &Longitude of orbital plane $p$ & $(180(p-1)/P)^\circ$\\
         \bottomrule
    \end{tabular}
    \label{tab:param}
\end{table}

The constellation is a Walker star, as illustrated in Fig.~\ref{fig:constellation}, with $P$ polar orbital planes deployed at a minimum altitude of $600$~km with orbital separation (i.e., altitude increments) of $10$~km. Each of these orbital planes consists of $N$ satellites. 

\begin{figure}[t]
    \centering
    \includegraphics{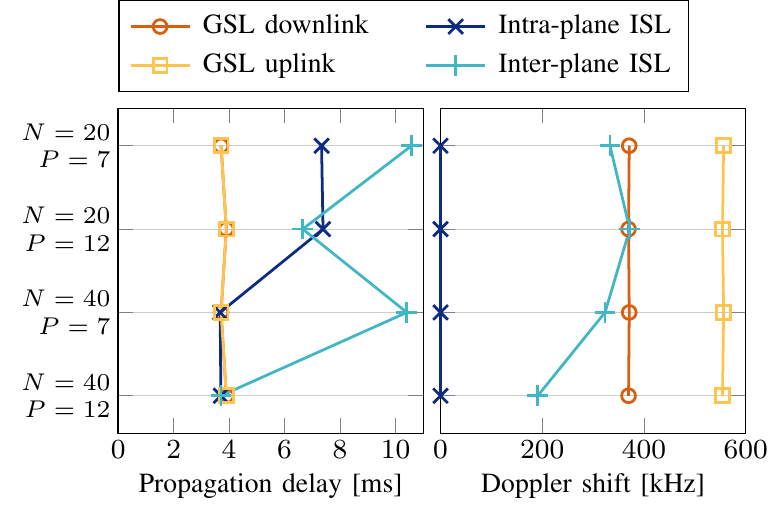}
    \caption{$95$th percentile of the propagation delay and Doppler shift at the physical links.}
    \label{fig:prop_delay_doppler}
\end{figure}

\begin{figure}[t]
    \centering
    \includegraphics{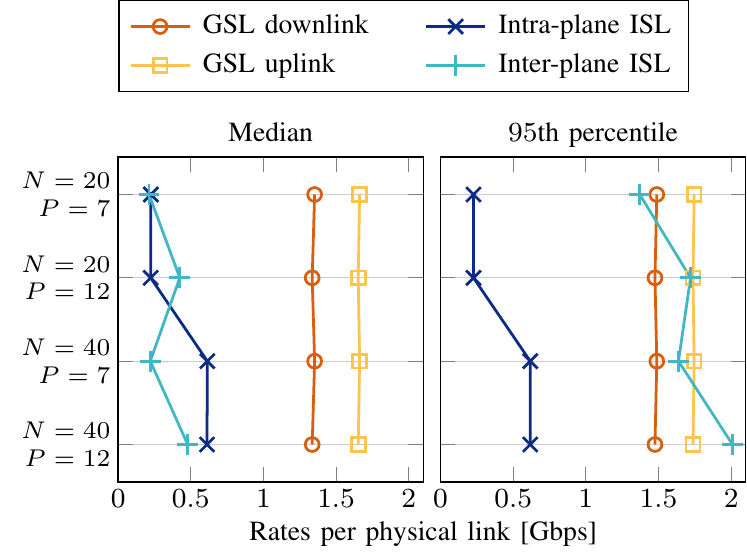}
    \caption{Median and $95$th percentile  of the data rates at the physical links in an interference-free environment.}
    \label{fig:rates}
\end{figure}

We consider a connection-oriented network, where the links are pre-established and the satellites and ground terminals have perfect beam steering capabilities. Hence, the gain at each established link is the maximum antenna gain. The inter-plane ISLs are established according to a greedy matching algorithm presented in~\cite{Soret2019} and thoroughly analyzed in~\cite{leyvamayorga2020}; the interested reader can find more technical details in this latter reference. The results presented in Fig.
~\ref{fig:prop_delay_doppler} and~\ref{fig:rates} for the GSL rates were obtained by distributing $10
^5$ users over the Earth's surface within the ground coverage of a satellite following a homogeneous Poisson point process (PPP).

Fig.~\ref{fig:prop_delay_doppler} shows the $95$th percentile of propagation delay and the Doppler shift for each physical link in the constellation. The latter is calculated as $f_D = v f_c/c$, where $v$ is the relative speed between the transmitter and the receiver, $f_c$ is the carrier frequency, and $c$ is the speed of light. Note that propagation delays of less than $4$~ms are typical in the GSLs. Besides, similar propagation delays were achieved at both the intra- and inter-plane ISLs with a total of $480$ satellites. Such short delays are physically unattainable in GEO systems and enable multi-hop transmissions that comply with the requirements established by the 3GPP for the user- and control-plane latency of $50$~ms round trip time (RTT)~\cite{3GPPTR38.913}. 

Time alignment is another challenge introduced by the long transmission distances. For example, the minimum propagation delay in the GSL to a satellite at $710$~km (i.e., the maximum altitude with $P=12$) is $2.3$~ms, which occurs at the zenith point. However, as illustrated in Fig.~\ref{fig:prop_delay_doppler}, the ground terminals near the edge of coverage of the same satellite will experience propagation delays that are up to $1.7$~ms longer. Hence, mechanisms are needed to accommodate or compensate for these temporal shifts.

Fig.~\ref{fig:prop_delay_doppler} also shows that a Doppler shift of nearly $600$~kHz is typical in the GSLs, which is comparable and even greater than that in the inter-plane ISLs. This is because cross-seam ISLs were not implemented. This is a common practice, for example, followed in the upcoming Kepler constellation, because of the huge Doppler shift when the satellites move in nearly opposite directions. For instance, if the cross-seam ISLs are implemented with $P=7$ and $N=20$, the $95$th percentile of the Doppler shift with is $1.46$~MHz. As a reference, the system bandwidth for NB-IoT is only $180$~kHz. In contrast, the intra-plane ISLs are not affected by Doppler shift because the intra-plane distances are rather stable. 

Next, Fig.~\ref{fig:rates} presents the median and $95$th percentile of the achievable instantaneous data rates at the physical links in an interference-free environment. That is, the rates are chosen from an infinite set of possible values to be equal to the capacity of an additive white Gaussian noise (AWGN) channel at specific time instants. 
Naturally, the distances in the GSLs and in the intra-plane ISLs are less variable than in the inter-plane ISLs. Because of this, the $95$th percentile of the rates is similar to the median in the GSLs and intra-plane ISLs but much greater for the inter-plane ISLs. 

Besides contributing to the Doppler shift, the movement of the satellites complicates the implementation of inter-plane ISLs by creating frequent and rapid changes in the inter-plane ISLs and greatly reducing the time a specific inter-plane ISL can be maintained, termed inter-plane \emph{contact times}. Hence, these links require frequent handovers, which involves neighbor discovery and selection (matching), as well as signaling for connection setup. Despite these challenges, implementing the inter-plane ISLs comes with massive benefits. For instance, Fig.~\ref{fig:rates} shows that the median of the achievable rates in the inter-plane ISLs are up to the par with those at the intra-plane ISLs. Moreover, the $95$th percentile of the rates at the inter-plane ISLs are close to those in the GSLs.

\subsection{Logical links}
\begin{figure}
    \centering
    \includegraphics{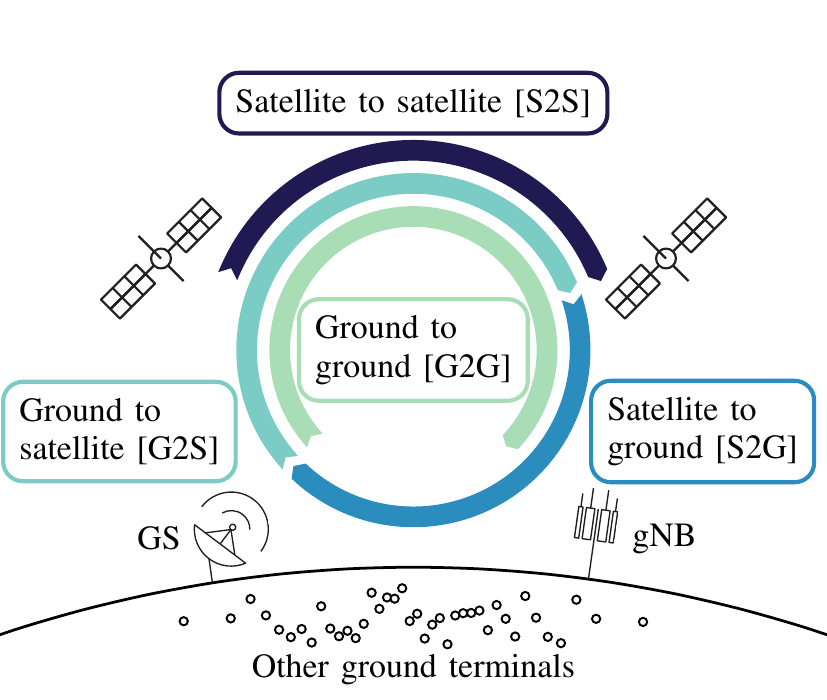}
    \caption{Sketch of the four logical links in a LEO constellation.}
    \label{fig:link_types}
\end{figure}

\emph{A logical link} is a path from the source transmitter to the end receiver. Hence, data travels over many different physical links, which may not be known by the two end-points. In our case, there are two different kinds of end-points, a satellite [S] and a ground terminal [G], which enables the definition of four logical links, which may utilize one or several of the physical links, GSL and ISL. These four logical links are illustrated in Fig.~\ref{fig:link_types} and described in the following.

\textbf{Ground to ground [G2G]:} Classical use of the network, where information is relayed between two distant points on the ground. Is also used for handover, routing and coordination of relays.

\textbf{Ground to satellite [G2S]:} Mainly used for maintenance and control operations initiated by the ground station. For example, to distribute instructions for ISL establishment and routing, but also for caching and telecontrol.

\textbf{Satellite to ground [S2G]:} Relevant when the satellites collect and transmit application data, such as in Earth observation, but also needed for handover and link establishment with GSs, radio resource management (RRM), fault detection, and telemetry.   

\textbf{Satellite to satellite [S2S]:} Mainly relevant for satellite-related control applications such as distributed processing, sensing, and routing, possibly exploiting distributed intelligence. Also used for topology management, including neighbor discovery, link establishment, and other autonomous operations in the space segment.

The use of the different physical and logical links is tightly related to the final application. In LEO constellations we identify not only applications that are inherited from terrestrial networks, but also those that are native to space communications. 

One exemplary application is the use of the constellation as a multi-hop relay network to increase the coverage of IoT deployments in rural or remote areas, where the cellular and other relaying networks are out of range~\cite{3GPPTR38.913}. In such applications, the IoT devices wake up periodically to send status updates. These updates are forwarded within the constellation until reaching the nearest satellite to the destination with an established link to a ground terminal. This end-to-end application relies on the [G2G] logical link.  

Another example are LEO constellations for Earth and/or space observation, both of which are native applications to satellite networks. In these, the satellites are equipped with cameras and sensors that can operate in the visible, near-infrared, thermal or microwave spectral domain. Naturally, the [S2G] is needed to retrieve the information in ground. Besides, the [S2S] link can be exploited for cooperation among satellites, for example, to point the cameras to a particular position when a first spacecraft detects an unusual event.

\section{Integration of LEO into 5G and beyond} \label{sec:5Gintegration}
 In this section we provide an overview of the status of the standardization process and the most relevant radio access technologies to support 5G services in LEO constellations. 

\subsection{3GPP: Ongoing Work}
Truly ubiquitous coverage is one of the major 5G drivers, and will only be possible through a close integration of satellite networks into 5G and B5G networks. For that, the 3GPP is working in the integration of Non-Terrestrial Networks (NTN) in future releases of 5G NR~\cite{ 3GPPTR22.822, 3GPPTR38.821}. This encompasses LEO, MEO, and GEO satellites, but also air-borne vehicles such as High Altitude Platforms (HAPs) operating typically at altitudes between $8$ and $50$~km. The goal is to ensure an end-to-end standard in the Release 17 timeframe -- the second phase of 5G -- originally scheduled for 2021. Specifically, a dedicated study for NTN IoT was agreed in December 2019, paving the way to introduce both narrowband~IoT (NB-IoT) and evolved MTC (eMTC) support with satellites. 


Two 5G satellite implementations are envisioned: transparent or regenerative payload~\cite{3GPPTR38.821}. In the first one, the satellites merely serve as relays toward the ground and, in the second one, satellites are a fully or partially functional gNBs that can perform, for example, encoding/decoding and routing. Hence, the regenerative payload implementation enables the use of the 5G logical interface between gNBs, the so-called Xn, to connect distant gNBs through the constellation. Moreover, the 3GPP considers two options of multi-connectivity in NTN, having the user equipment (UE) connected to one satellite and one terrestrial network, or to two satellites~\cite{3GPPTR38.821}. Another interesting application of LEO constellations is to backhaul fixed or moving terrestrial gNBs in areas with no additional terrestrial infrastructure. 

There are two options to connect the 5G UEs to the constellation. The first option is through a gateway (i.e., a relay node), which uses the constellation for backhaul. The big advantage of this approach is that legacy UEs are fully supported and no additional RF chain is required. The second option is having the UEs to communicate directly with the satellite or the HAP. With this second option, the coverage of the constellation is maximized, but the limited transmission range of the UEs becomes the main challenge. Nevertheless, this may not be a hindrance in the case of low-power wide-area technologies such as NB-IoT in selected scenarios: the maximum NB-IoT coupling loss of $164$~dB is, in principle, sufficient to directly communicate at $2$~GHz (i.e., S-band) with LEO satellites at heights up to $700$~km. Specifically, with these parameters, the maximum FSPL for an elevation angle of $30^\circ$ is $163.31$~dB.

The identification of the three 5G services is done in NR with the standardized 5G Quality of Service (QoS) Identifiers (5QIs). These consist of a set of pre-set values for the most frequently used services, such as the latency budget, the maximum error rate or the priority levels. Moreover, dynamic 5QIs can be defined for services that do not fit the pre-defined list. With the 5QI the network is able to properly treat the packets according to the type of service. To account for the fundamental limitations and/or performance differences between terrestrial and satellite networks, it has been proposed to include new Radio Access Technology (RAT) identifiers to indicate in each packet that a UE is using an NTN. This is necessary to, e.g., avoid a timer expiration if the 5QI is not compatible with the RAT type~\cite{3GPPTR23.737}.

\subsection{Physical layer}
The waveform defines the physical shape of the signal that carries the modulated information through a channel. In NR, the defined waveform is based on Orthogonal Frequency Division Multiplexing (OFDM)~\cite{Kodheli2017,Guidotti2019}, which is very sensitive to Doppler shifts. As observed in Section~\ref{sec:phylinks}, accurate Doppler compensation and subcarrier spacing must be put in place to tolerate Doppler shifts of up to $600$~kHz in the GSLs. 

To overcome these limitations, several alternatives have been intensively studied in the literature over the past few years, such as Universal Filtered Multi-Carrier (UFMC), Generalized Frequency Division Multiplexing (GFDM) and Filter Bank Multi-Carrier (FBMC)~\cite{Wunder2014}. These waveforms allow for higher robustness against Doppler shifts and flexible time-frequency resource allocation in exchange for a higher equalization complexity \antcom{regardless of the magnitude of the Doppler shift. However, in case of severe Doppler shifts, Factor Graph based equalization for FBMC transmissions outperforms the OFDM system in terms of complexity and performance \cite{Woltering2018}.} Nevertheless, maintaining OFDM is convenient to provide compatibility with terrestrial UEs.

Modulation schemes for satellite communications usually involve low modulation orders for robustness and low peak-to-average power ratio (PAPR) to enable the use of nonlinear power amplifiers. The preferred choice in recent commercial LEO missions is amplitude and phase-shift keying (APSK) with modulation orders up to $16$~\cite{Portillo2019}. Hence, the most suitable modulation schemes supported by 5G NR are quadrature phase-shift keying (QPSK) and quadrature amplitude modulation (QAM) with modulation order $16$.

Terrestrial gNBs adapt the modulation and coding scheme to the current channel conditions, for which the UEs must transmit information about the channel quality to the gNB~\cite{Guidotti2019}. In satellite systems, the channel conditions are mainly determined by the path loss, which can be easily predicted from the constellation geometry. In addition, rain fade, antenna pointing errors, noise, and interference create frequent, yet minor, changes.

Nevertheless, the high orbital velocities of LEO satellites creates significant, yet predictable, changes in the path loss at the GSL.
For the transmission of short packets in the GSL, the priority is to minimize link outages and packet errors to avoid long RTTs for feedback. In these cases, fixed robust modulation and coding schemes are preferred.

\begin{figure}[t]
    \centering
    \includegraphics[width=\columnwidth]{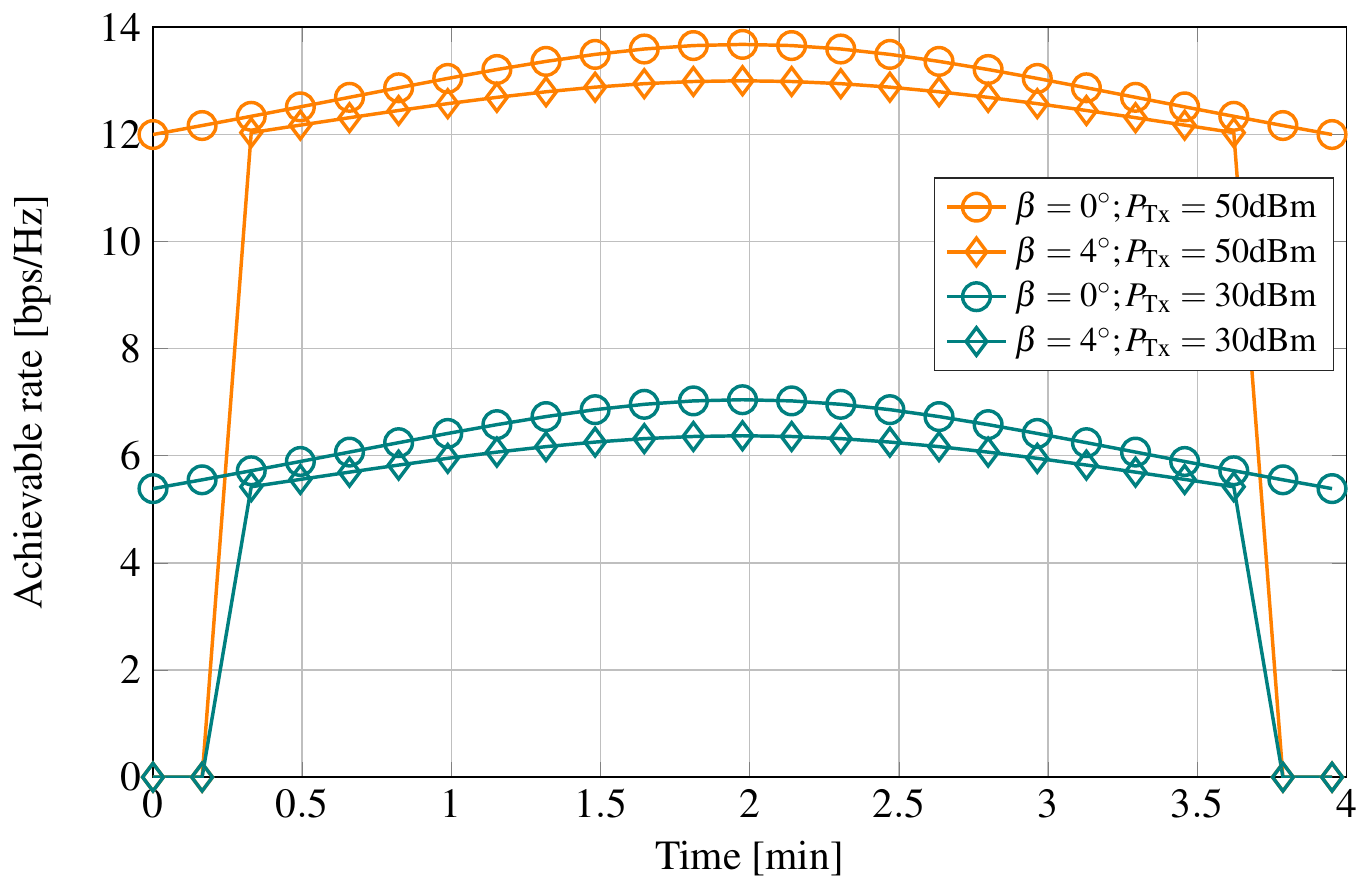}
    \caption{Evolution of \antcom{achievable rate} for GSL downlink during one pass for ground terminals with a longitude shift $\beta\in\{0^\circ,4^\circ\}$ and transmission power $P_\text{Tx}\in\{30,50\}$\,dBm.} 
    \label{fig:capacity_time}
\end{figure}

On the other hand, adaptive coding and modulation is interesting for long packet transmissions in the GSL, where the predictability of the path loss can be exploited. As an example, Fig.~\ref{fig:capacity_time} shows the evolution of the achievable \antcom{rate} with time for a LEO satellite in a polar orbit at an altitude of $600$~km for two different ground terminals deployed along the Equator. The first is the optimal pass, where the shift in longitude between the ground terminal and satellite is $\beta=0$ and, the second one, is a typical pass where $\beta=4^\circ$. Naturally, communication is not possible when the elevation angle is below the minimum, thus, the \antcom{achievable rate} becomes zero for that case. The simulation parameters, except the transmission power, are those listed in Table
~\ref{tab:param}.

As it can be seen, the peak of the \antcom{achievable rate} occurs at around $2$~minutes after the satellite establishes the GSL with the ground terminal at $\beta=0^\circ$. This is because the duration of the optimal pass is $4.1$~minutes in this example. At this peak, the \antcom{achievable rate} can be up to \antcom{$31$\% and $14$\%} higher than the minimum for $P_\text{Tx}=30$~dBm and $P_\text{Tx}=50$~dBm, respectively for the ideal pass. In comparison, the pass of the ground terminal at $\beta=4^\circ$ is around $0.8$~minutes shorter and its peak \antcom{achievable rate} is around $1$~bps/Hz lower for both considered power levels, which is significant. 

\antcom{
MIMO is yet another key enabler for high data rate communications. Although MIMO has tremendous potential for increased spectral efficiency in eMBB traffic through LEO constellations, MIMO techniques have, so far, been mostly restricted to terrestrial communication systems. Pure LoS connections and the large transmission distances between satellites and ground terminals introduce some unique challenges that need to be overcome. In particular, exploiting the full MIMO gain requires a large array aperture, i.e., large distances between transmit and/or receive antennas~\cite{Schwarz2019}. In the space segment, this separation can be implemented by cooperatively transmitting to a GS from multiple satellites, flying in close formation}.

\begin{figure}[t]
    \centering
    \includegraphics[width=\columnwidth]{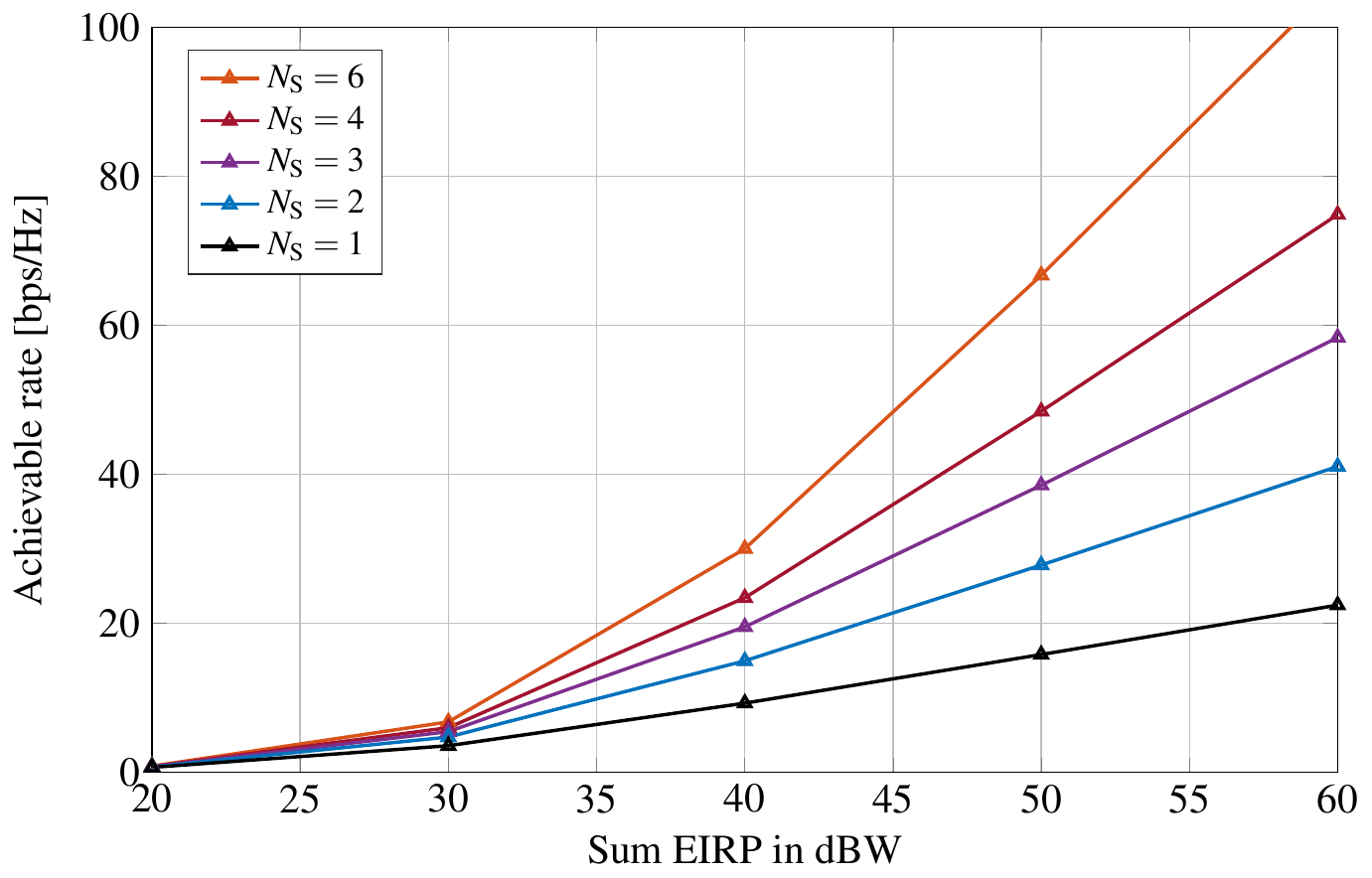}
    \caption{\antcom{Achievable rate} for $N_S$ satellites in close formation and simultaneously transmitting towards a GS.}
    \label{fig:capacity_swarm}
\end{figure}

The expected gain of such a setup is evaluated in Fig.~\ref{fig:capacity_swarm}, where the \antcom{achievable rate} of $N_S=\{1,2\dotsc,6\}$ satellites transmitting cooperatively to a single GS as a function of the sum EIRP of all $N_\text{S}$ satellites is shown.
The satellites are assumed to fly in a dense trail formation in the same orbital plane at an altitude of $600\,$km with an inter-satellite distance of
$100$\,km. Each of these satellites is equipped with $N_t=12/N_\text{S}$ antennas, such that the total number of transmit antennas is $12$ for all cases.
The GS is equipped with a uniform linear array (ULA) consisting of $100$ antennas having a gain of $G_\text{Rx}=20\,$dBi each, e.g., planar antennas with $100\times100$ elements. The ULAs axis is aligned with the ground trace of the satellites.
The downlink transmission takes place in the Ka-Band at a carrier frequency of $f_c=20\,$GHz (as listed in Table~\ref{tab:param}) and GS antennas are spaced $\lambda_c/2 = 7.5\,\text{mm}$ apart.

It can be observed that the \antcom{achievable rate} increases with the number of transmitting satellites, although neither the total transmit power nor the total number of antennas increases. Hence, transmitting cooperatively from several satellites provides a considerable gain over equipping a single satellite with a large antenna array. In addition, this joint transmission allows to form very narrow beams which leads to better spatial separation and, thus, higher spectral efficiency when serving different GSs located geographically close to each other on the same time-frequency resources~\cite{Roper2019}.

\subsection{Radio access}

\emph{Radio access} in the GSL is, essentially, random access (RA) due to the large amount of nodes and the fact that both the number of ground terminals and the traffic patterns are not known in advance. 
\pp{There are two principal types of RA protocols:} grant-based and grant-free. Grant-based RA is the go-to solution in 5G and the newly proposed new two-step random access procedure can mitigate the excessive delay and time alignment issues of the legacy four-step procedure in satellite communications~\cite{3GPPTR38.821}. Nevertheless, due to the large individual coverage of the satellites, the capacity of the grant-based RA of 5G can be easily surpassed if the ground terminals are allowed to perform direct access to the satellites.

On the other hand, grant-free RA is preferred for the transmission of short and infrequent data packets that characterizes massive IoT. Nevertheless, the long distance between end points prevents the use of traditional channel sensing protocols~\cite{Cioni2018}. Instead, non-orthogonal medium access (NOMA) techniques that incorporate successive interference cancellation (SIC) may be better suited~\cite{DeGaudenzi2018}. 

Due to space, weight and budget limitations, small-satellite may incorporate relatively wide-beam antennas~\cite{Budianu2013}. Therefore, interference mitigation in the ISL may be required.
In the intra-plane ISL, the transmitter and the receiver do not change because the relative distances are preserved. Therefore, fixed access schemes like Frequency Division Multiple Access (FDMA) or Code Division Multiple Access (CDMA) are simple and attractive solutions~\cite{Rahakrishnan2016}. With FDMA, the frequency reuse factor to mitigate interference along the orbit must be properly designed, which comes at the cost of higher bandwidth requirements. On the other side, the challenges of CDMA, for example, synchronization or near-far effects, can be overcome by using asynchronous CDMA with non-orthogonal codes. 

On the other hand, in the inter-plane ISL, the satellite pairs may change frequently over time and the distances are not preserved. This is specially true because commercial LEO constellations usually present slight asymmetries in the orbital planes to minimize the risk of (physical) collisions between satellites~\cite{Lewis2019}.
 Nevertheless, the predictability of the constellation geometry can still be exploited by a centralized entity (e.g., a GS or a satellite with sufficient storage space and processing power) to allocate orthogonal resources for inter-plane communication. By doing so, an efficient resource utilization can be achieved while mitigating the potential interference among inter-plane, but also between inter-plane and intra-plane ISL.

\begin{figure}
    \centering
    \includegraphics{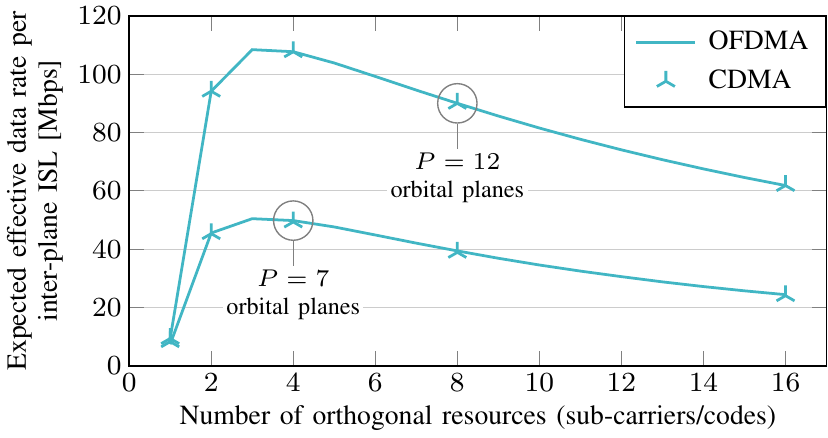}
    \caption{Effective data rates at the inter-plane ISLs of a Walker star constellation with OFDMA and CDMA.}
    \label{fig:rates_ma}
\end{figure}

Fig.~\ref{fig:rates_ma} illustrates the expected data rate per inter-plane ISL in the Walker star constellation (see Fig.~\ref{fig:constellation}) with $N=20$. The allocation of orthogonal resources takes place with a greedy algorithm, described in~\cite{leyvamayorga2020}, which makes the best global choice at each iteration. The parameters settings are listed in Table~\ref{tab:param}. A worst-case scenario for interference is considered (i.e., with omnidirectional antennas). The data rates are calculated to ensure zero outage probability and the interference is treated as AWGN. The cross-seam ISLs are not implemented due to the large Doppler shift.

Two types of multiple access methods are considered: 1) orthogonal FDMA (OFDMA), where the bandwidth $B$ is divided into a given number of sub-carriers of the same bandwidth $B_\text{sc}$, and 2) CDMA, where $K\in\{2^k:k\in\mathbb{Z}\}$ orthogonal Walsh codes, whose spreading factor is equal to $K$, are used. The data rates with both OFDMA and CDMA were calculated to ensure a zero-outage probability with the maximum feasible value of the interference for all the established ISLs. Note that, for OFDMA, the EIRP decreases as the number of orthogonal subcarriers increases due to the constant EIRP density.

As Fig.~\ref{fig:rates_ma} shows, the number of orthogonal resources to achieve the peak rates is relatively small: $3$ with OFDMA and $4$ with CDMA (since $K=3$ is not a feasible value). Besides, note that both OFDMA and CDMA lead to closely similar rates for all feasible values of $K$. The reason for this is that, for OFDMA, the decrease in noise power with narrow sub-carrier bandwidths is compensated by a decrease in EIRP. On the other hand, the rates with $P=12$ are usually more than $2\times$ greater than with $P=7$. This non-linear increase of the data rates with $N$ was also observed in Fig.~\ref{fig:rates}.

The results presented in Fig.~\ref{fig:prop_delay_doppler}, \ref{fig:rates}, and~\ref{fig:rates_ma} were obtained considering that the inter-plane ISL are established with the single objective of maximizing the sum of rates in the LEO constellation. However,
 there are situations in which multiple satellites can benefit from establishing an inter-plane connection with a specific satellite at the same time. In these cases, the inter-plane ISL can be seen as a mesh network. Unlike terrestrial mobile ad-hoc networks, the position of the satellite neighbors can be predicted if the orbital information is available at each node. Protocols from mesh networks for connecting directly, dynamically and non-hierarchically to as many other satellites as possible and cooperate with one another can, therefore, be adapted to the specific conditions of the LEO constellation. For instance, the use of relatively wide beams enables the use of distributed approaches, such as the deferred acceptance algorithm~\cite{Gu2015}, where the inter-plane ISL may be established opportunistically based on the individual preferences of the satellites.

\subsection{Radio Slicing}
A general-purpose satellite constellation must support the heterogeneity of eMBB, URC, and mMTC services. Besides, the user, control and TMTC traffic have widely different characteristics and requirements. 
\emph{Network slicing}~\cite{Ferrus2018} 
is a key 5G feature to support heterogeneous services and to provide performance guarantees by avoiding performance degradation due to other services. In the Radio Access Network (RAN), the conventional approach to \emph{radio slicing} is to allocate orthogonal radio resources at the expense of a lower network efficiency. Instead, non-orthogonal slicing may bring benefits in terms of resource utilization at the expense of a reduced predictability in the QoS. In particular, non-orthogonal slicing in the RAN, in the form of NOMA for heterogeneous services, may lead to better performance trade-offs than orthogonal slicing in terrestrial communications~\cite{Popovski2018}. However, extensions to these communication theoretic models are needed to characterize the potential gains of non-orthogonal slicing in LEO small-satellite constellations.

The time and frequency multiplexing of services and data traffic with widely different characteristics and requirements introduces similar challenges as in terrestrial networks. For instance, it requires priority-aware mechanisms in the data link and medium access layers to guarantee the efficient delivery of critical packets. However, the space multiplexing has some advantages in LEO constellations, thanks to the line-of-sight conditions, the link diversity, and the availability of multiple antennas with narrow beams.

Until now, LEO, MEO and GEO have been addressed separately in 3GPP. Nevertheless, hybrid architectures combining different orbits may play a major role in future networks~\cite{Chien2019} and contribute to the network slicing. Hybrid solutions are reminiscent of the evolution towards heterogeneous cellular networks and the mix of cell sizes, but with the added advantage that the diversity in orbits and satellite capabilities can complement each other favorably. For instance, the short transmission times between ground terminals and LEO satellites can be combined with: 1) the wide coverage and the great communication and computation capabilities of GEO satellites and 2) with the available navigation satellites at MEO (e.g., GPS, Glonass, and Galileo constellations). Hence, hybrid architectures provide great flexibility and an increased capacity to accommodate a heterogeneity of application requirements.


\section{Conclusions} \label{sec:conclusions}
In this paper, we described the main opportunities and connectivity challenges of LEO small-satellite constellations. Besides, we characterized the physical links in LEO constellations in terms of propagation delay, Doppler shift, and achievable data rates. Our results showcase that LEO constellations have the potential to fulfill the 5G promise of true ubiquity and to support the generic use cases of eMBB and mMTC, as well as URC with latency requirements of a few tens of milliseconds. Nevertheless, these physical links present different characteristics and, hence, must be properly designed to unleash the full potential of the constellation.
 Furthermore, we provided an overview and taxonomy for the logical links, in connection with the used physical links and relevant use cases. Finally, we discussed about several PHY/MAC and radio slicing enabling technologies and outlined their role in supporting 5G connectivity through LEO satellites. Our results illustrate the potential benefits of adaptive coding and modulation schemes in the GSL and of proper resource allocation for multiple access in the ISL.
\bibliographystyle{IEEEtran}
\bibliography{main}

\begin{IEEEbiography}[{\includegraphics[width=1in,height=1.25in,clip,keepaspectratio]{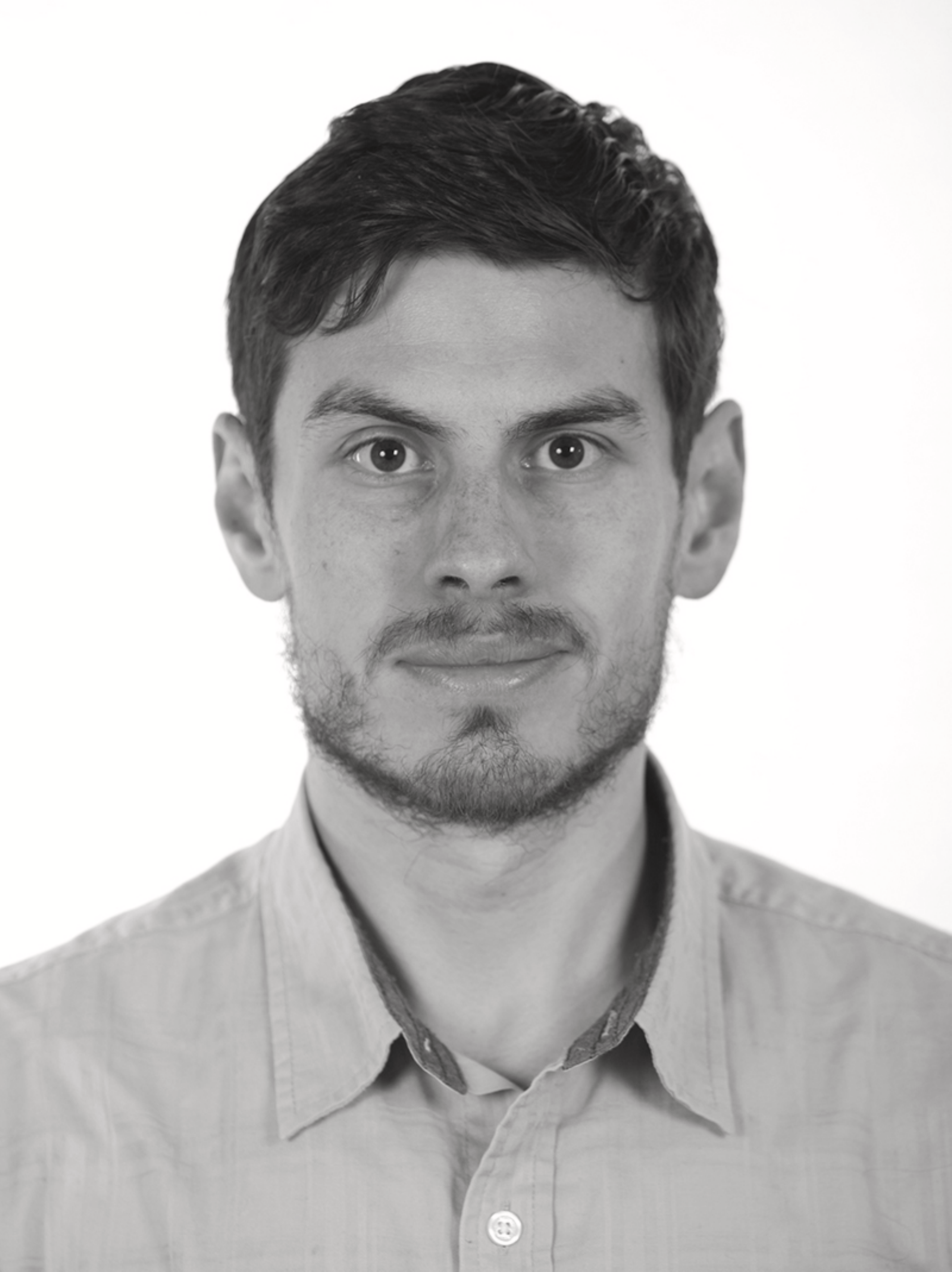}}]{Israel Leyva-Mayorga} (M'20) received the M.Sc. degree (Hons.) in mobile computing systems from the Ins\-ti\-tu\-to Po\-li\-t\'ec\-ni\-co Na\-cio\-nal, Mexico, in 2014 and the Ph.D. (\emph{Cum Laude}) in telecommunications from the U\-ni\-ver\-si\-tat Po\-li\-t\`ec\-ni\-ca de Va\-l\`en\-cia, Spain, in 2018. He was a Visiting Researcher at the Department of Communications, UPV, in 2014, and at the Deutsche Telekom Chair of Communication Networks, Technische Universit\"at Dresden, Germany, in 2018. He is currently a Postdoctoral Researcher at the Connectivity Section (CNT) of the Department of Electronic Systems, Aalborg University, Denmark. His research interests include 5G and beyond, satellite networks, random access protocols, and massive machine-type communications.
\end{IEEEbiography}

\begin{IEEEbiography}[{\includegraphics[width=1in,height=1.25in,clip,keepaspectratio]{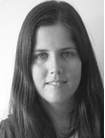}}]{Beatriz Soret} (M'11) received the M.Sc. and Ph.D. degree in Telecommunications from the Universidad de Malaga (Spain), in 2002 and 2010, respectively. She is currently an associate professor at the Department of Electronic Systems, Aalborg University (Denmark). Before, she has been with Nokia Bell-Labs and GomSpace. She has co-authored more than 60 publications in journals and conference proceedings, and 16 patents in the area of wireless communications. Her research interests are within satellite communications with LEO constellations, low-latency and high reliable communications, and 5G and post-5G systems. 
\end{IEEEbiography}

\begin{IEEEbiography}[{\includegraphics[width=1in,height=1.25in,clip,keepaspectratio]{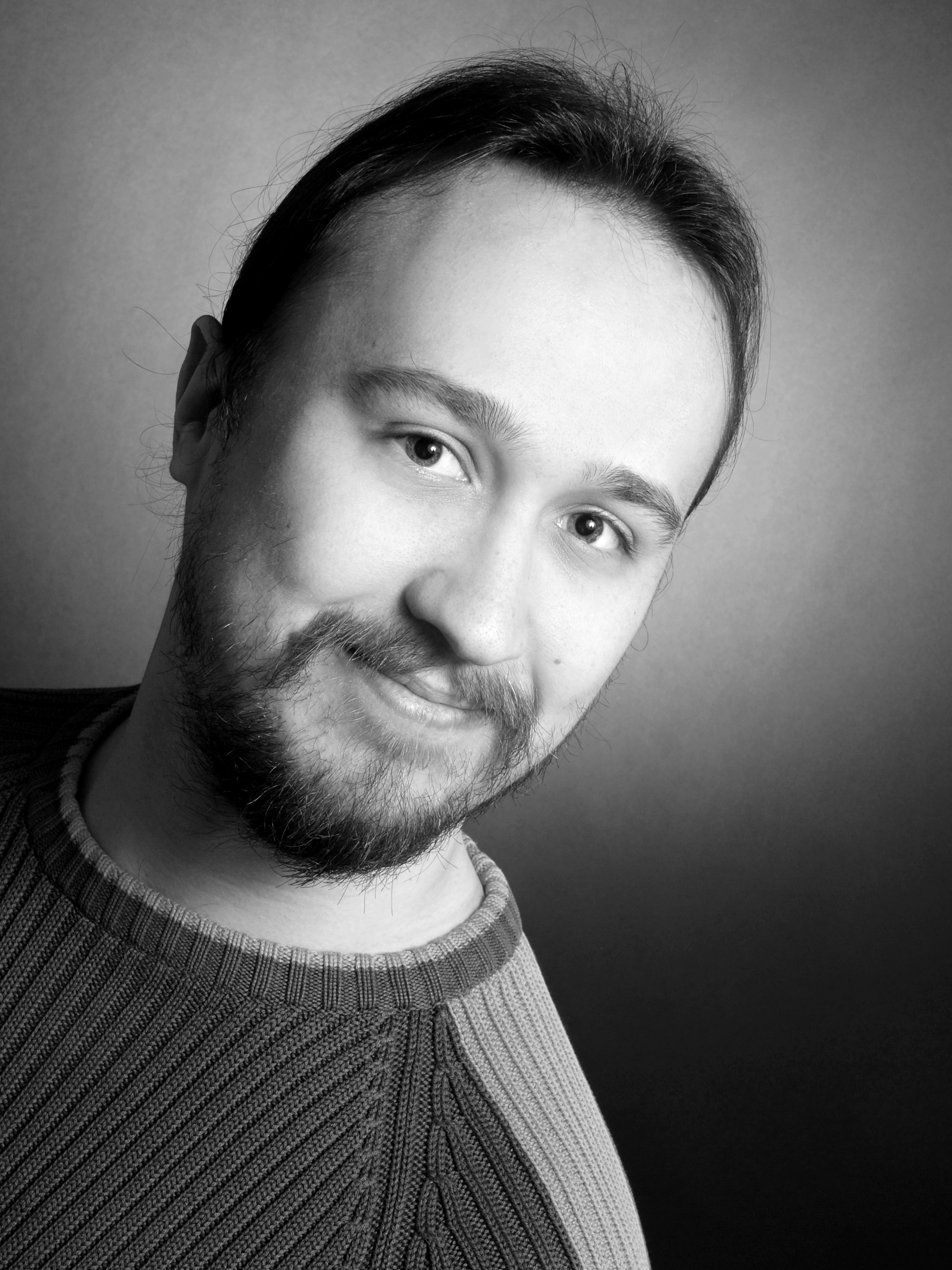}}]{Maik R{\"o}per}
(S'18) received the B.Sc. and M.Sc. degree in electrical engineering and information technology from the University of Bremen, Germany, in 2014 and 2016, respectively. Since then, he has been a research assistant with the Department of Communications Engineering at the University of Bremen, where he is currently pursuing the Ph.D. degree. His research interests include satellite communications, precoding and distributed signal processing.
\end{IEEEbiography}

\begin{IEEEbiography}[{\includegraphics[width=1in,height=1.25in,clip,keepaspectratio]{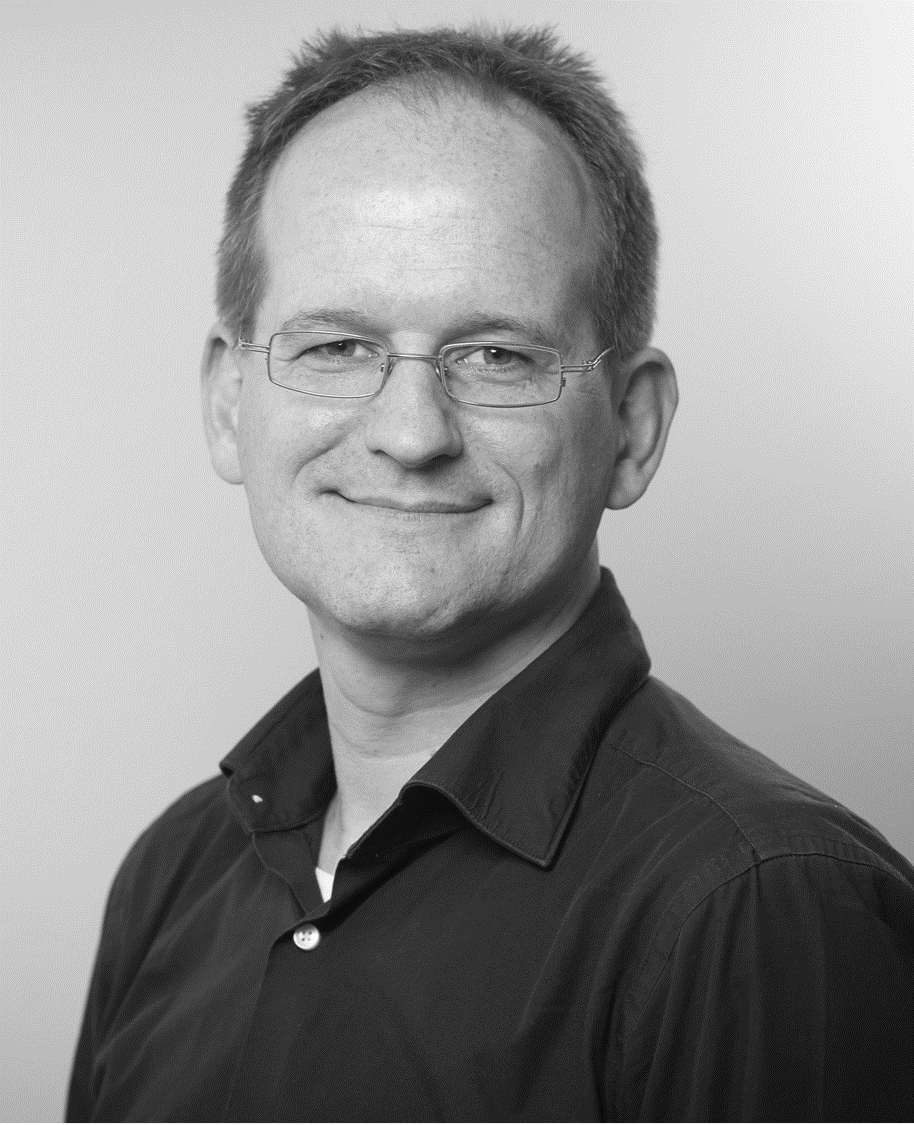}}]{Dirk W{\"u}bben}
(S'01--M'06--SM'12) is a senior researcher group leader and lecturer at the Department of Communications Engineering, University of Bremen, Germany. He received the Dipl.-Ing. (FH) degree in electrical engineering from the University of Applied Science M\"{u}nster, Germany, in 1998, and the Dipl.-Ing. (Uni) degree and the Dr.-Ing. degree in electrical engineering from the University of Bremen, Germany in 2000 and 2005, respectively. His research interests include wireless communications, signal processing, multiple antenna systems, cooperative communication systems, channel coding, information theory, and machine learning. He has published more than 140 papers in international journals and conference proceedings. He has been an Editor of IEEE Wireless Communication Letters. He is a board member of the Germany Chapter of the IEEE Information Theory Society and member of VDE/ITG Expert Committee ``Information and System Theory''.
\end{IEEEbiography}

\begin{IEEEbiography}[{\includegraphics[width=1in,height=1.25in,clip,keepaspectratio]{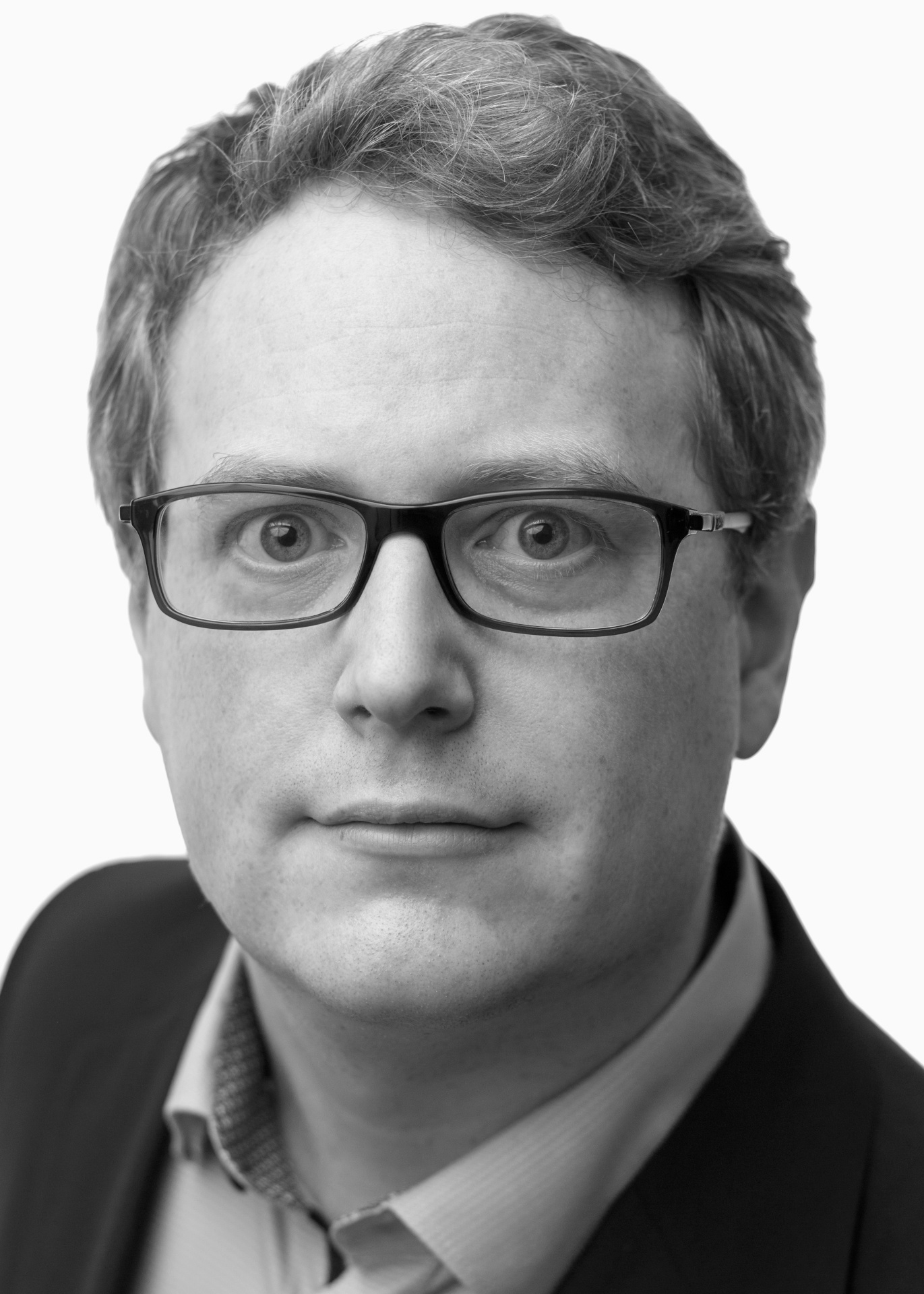}}]{Bho Matthiesen}
(S'12--M'20) received the Diplom-Ingenieur (M.Sc.) degree and his Ph.D.~(with distinction) from Technische Universit\"at Dresden, Germany, in 2012 and 2019, respectively. Since 2020, he is a research group leader at the U Bremen Excellence Chair in the Department of Communications Engineering, University of Bremen, Germany. He serves as publication chair at ISWCS 2021 and is an Associate Editor for EURASIP Journal on Wireless Communications and Networking. His research interests are in communication theory, wireless communications, and optimization theory.
\end{IEEEbiography}

\begin{IEEEbiography}[{\includegraphics[width=1in,height=1.25in,clip,keepaspectratio]{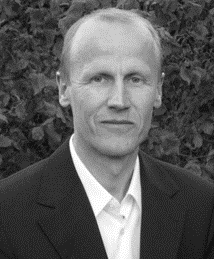}}]{Armin Dekorsy}(SM'18) received the B.Sc. degree from Fachhochschule Konstanz, Germany, the M.Sc. degree from the University of Paderborn, Germany, and the Ph.D. degree from the University of Bremen, Germany, all in communications engineering. From 2000 to 2007, he was a Research Engineer with Deutsche Telekom AG and a Distinguished Member of Technical Staff at Bell Labs Europe, Lucent Technologies. In 2007, he joined Qualcomm GmbH as a European Research Coordinator, conducting Qualcomms internal and external European research activities. He is currently the Head of the Department of Communications Engineering, University of Bremen. Prof. Dekorsy is a senior member of the IEEE Communications and Signal Processing Society and the head of VDE/ITG Expert Committee ``Information and System Theory.'' He has authored or coauthored more than 180 journal and conference publications. He holds more than 19 patents in the area of wireless communications. Prof. Dekorsy investigates new lines of research in wireless communications and signal processing for transmitter baseband design which can readily be transferred to industry. His current research directions are distributed signal processing, compressive sensing, and machine learning.
\end{IEEEbiography}

\begin{IEEEbiography}[{\includegraphics[width=1in,height=1.25in,clip,keepaspectratio]{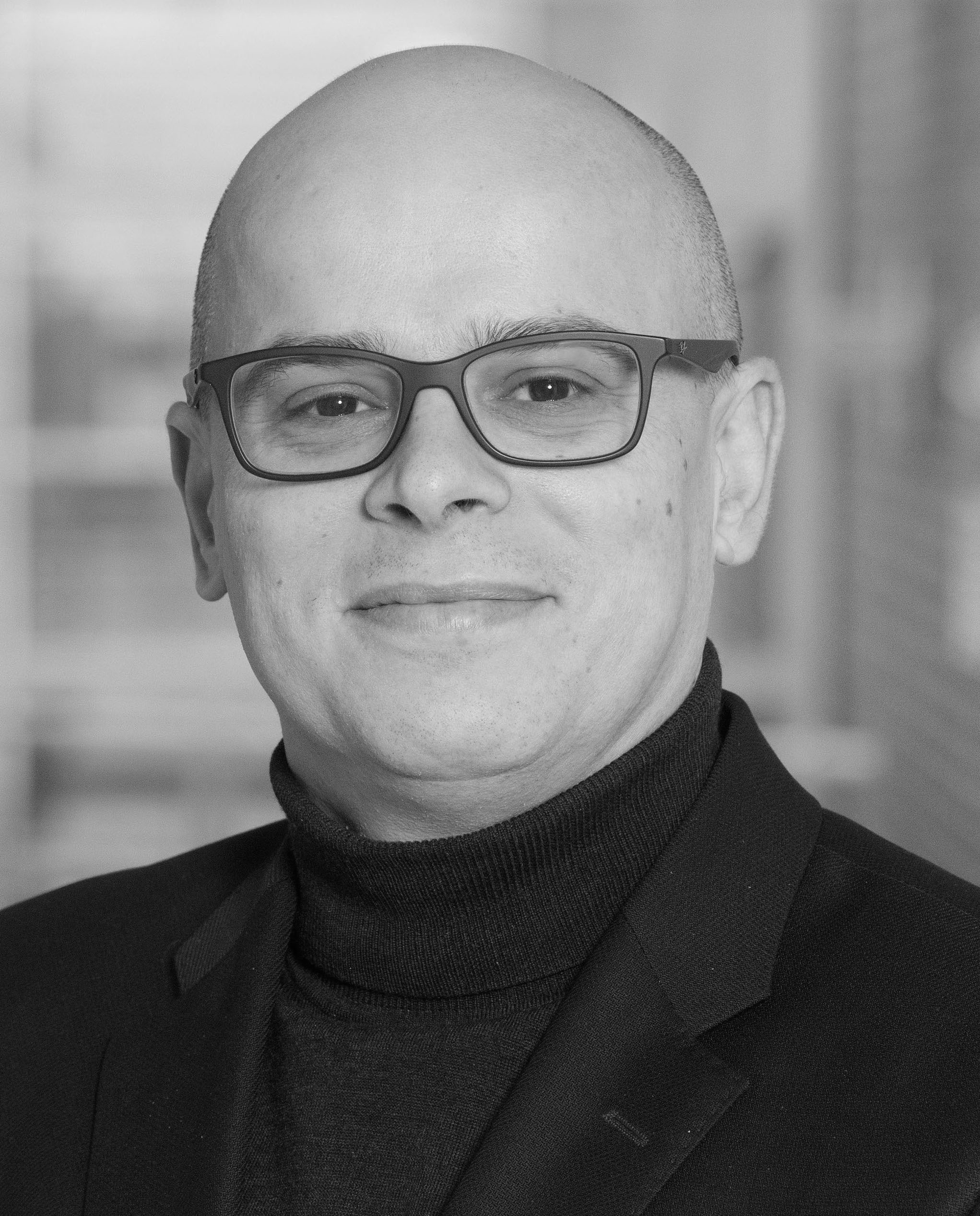}}]{Petar Popovski} (S'97--A'98--M'04--SM'10--F'16) is a professor at Aalborg University, where he heads the Connectivity Section, and holder of a U Bremen Excellence Chair at the University of Bremen. He received his Dipl.-Ing./Magister Ing. in communication engineering from Sts. Cyril and Methodius University in Skopje and his Ph.D. from Aalborg University. He received an ERC Consolidator Grant (2015) and the Danish Elite Researcher award (2016). He is an Area Editor for IEEE Transactions on Wireless Communications. Since 2019, he is a Member-at-Large of the Board of Governors of the IEEE Communications Society. His research interests are in the area of wireless communication, communication theory and Internet of Things. In 2020 he published the book ``Wireless Connectivity: An Intuitive and Fundamental Guide''. 
\end{IEEEbiography}

\EOD

\end{document}